\documentclass[aps,pra,reprint,superscriptaddress]{revtex4-2}

\usepackage{graphicx}
\usepackage{dcolumn}
\usepackage{bm}
\usepackage{amsmath,amssymb}
\usepackage{amsthm}
\usepackage{tikz}
\usepackage{braket}
\usepackage{quantikz}
\usepackage{xcolor}
\usepackage{url}

\usepackage{hyperref}
\usepackage{color}

\begin{document}

\title{Time-of-Flow Distributions in Discrete Quantum Systems: From Operational Protocols to Quantum Speed Limits}

\author{Mathieu Beau}
\affiliation{Department of Physics, University of Massachusetts, Boston, Massachusetts 02125, USA}

\date{\today}

\begin{abstract}
We propose a general and experimentally accessible framework to quantify 
transition timing in discrete quantum systems via the time-of-flow (TF) distribution. 
Defined from the rate of population change in a target state, the TF distribution can be 
reconstructed through repeated projective measurements at discrete times on 
independently prepared systems, thus avoiding Zeno inhibition. In monotonic regimes, it 
admits a clear interpretation as a time-of-arrival (TOA) or time-of-departure (TOD) 
distribution. We apply this approach to optimize time-dependent Hamiltonians, analyze 
shortcut-to-adiabaticity (STA) protocols, study non-adiabatic features in the dynamics of a three-level time-dependent detuning 
model, 
and derive a transition-based quantum speed limit (TF-QSL) for both closed and open 
quantum systems. We also establish a lower bound on temporal uncertainty and examine 
decoherence effects, demonstrating the versatility of the TF framework for quantum 
control and diagnostics. This method provides both a conceptual tool and an experimental 
protocol for probing and engineering quantum dynamics in discrete-state platforms.
\end{abstract}

\maketitle

\section{Introduction}

Time plays a peculiar role in quantum mechanics: while space is associated with self-adjoint operators and measurable observables, time remains an external parameter. This fundamental asymmetry leads to challenges in defining meaningful \textit{{time-of-arrival}} (TOA) distributions, especially when compared to the clear probabilistic framework for position measurements~\cite{pauli1933handbuch}. Numerous approaches have been proposed for continuous systems (see reviews~\cite{Muga1,Muga2}), yet a consensus remains elusive due to issues such as ambiguity in measurement interpretation, non-uniqueness, and~divergence in limiting cases. In~contrast, the~case of \textit{discrete quantum systems}, relevant to qubit-based platforms, spin chains, and~trapped ions, has received far less attention. Proposals such as the Page and Wootters’ mechanism~\cite{Wootters83} or the quantum clock approach,~\cite{Giovannetti15,Maccone20}, where time is inferred from correlations between a system and an auxiliary quantum clock, are, in~principle, applicable to discrete systems. Despite their conceptual depth and promising implications, none of these proposals have yet been explored experimentally in the laboratory. 
Beyond being a purely conceptual problem, the~notion of time is also crucial in modern quantum technologies, in~particular for optimal control and decoherence, which both have great importance in quantum computing~\cite{Gambetta08}, quantum control~\cite{WisemanMilburn2010}, quantum simulation~\cite{Cirac12}, quantum metrology~\cite{Maccone03,Maccone12,Tóth_2014}, shortcut to adiabaticity (STA) \cite{Chen10,An16,delCampo19}, and~finite-time quantum thermodynamics~\cite{Watanabe17}.

Different approaches offer a variety of methods for the quantification of time, which can be used to help quantum technology speed up processes. 
Quantum speed limits (QSLs) \cite{mandelstam1945uncertainty,margolus1998maximum} 
are useful concepts that tell us how fast a transition could happen and have been used for 
optimal control~\cite{Deffner17},~decoherence~\cite{delCampo13}, and~quantum 
metrology~\cite{Maccone03,Maccone20}. However, while QSLs offer lower bounds on 
transition times, they do not provide experimentally accessible information about the time 
distribution of the transition. 
Transition timing in discrete quantum systems has been approached through quantum trajectories~\cite{Gambetta08}, which model dynamics under continuous monitoring by unraveling the evolution into stochastic jumps or diffusive paths. While this framework captures decoherence and measurement backaction, it requires open-system modeling and continuous observations, which can induce a quantum Zeno effect. Alternatively, first-detection protocols~\cite{Liu20} define transition times as the earliest successful projective measurement following stroboscopic sampling. This approach yields useful timing statistics but focuses on first-event detection and relies on invasive monitoring or collapse-based dynamics, limiting their generality and experimental accessibility in closed or control-oriented~settings.

As seen above, the~passage of time in discrete quantum systems lacks a unifying 
framework that addresses both theoretical and experimental challenges. Such a framework 
should not only clarify conceptual issues, such as the time-of-arrival problem, but~also 
support practical, yet fundamental, tasks in quantum engineering. In~this work, we 
propose a general and operationally meaningful concept of \textit{time-of-flow (TF) 
distribution} to bridge the gap between these complementary aspects. The~TF distribution 
is defined from the absolute value of the time derivative of the probability to occupy a 
target state and~is experimentally reconstructible via projective measurements at discrete 
time steps on independently prepared systems, thus \textit{bypassing the quantum Zeno 
effect} for continuous measurement protocols~\cite{Wineland90}. In~monotonic regimes, 
this TF distribution is interpreted as the time-of-arrival (TOA) or time-of-departure (TOD) 
profiles and yields expected results in limiting cases (delta pulse model).
Beyond conceptual consistency, we demonstrate how the TF distribution provides a practical framework to extract meaningful timing statistics. We apply this approach to optimize shortcuts to adiabaticity (STA)
, and~to analyze \textit{decoherence dynamics} in open quantum systems described by Lindblad-type master equations. This framework, therefore, offers both a theoretical tool and an experimental protocol to explore quantum timing in discrete-state systems. 
In addition, we derive a TF-based quantum speed limit (TF-QSL) and a new uncertainty relation linking the temporal spread of transitions to their energetic or dynamical generators. These results position the TF distribution as both a practical diagnostic and a fundamental tool for quantum dynamics in discrete~systems.

\section{Heuristic, Empirical and Theoretical Definition of the Time-of-Flow for Discrete Quantum~Systems}

Heuristically, the~time-of-flow (TF) distribution quantifies \emph{how the probability of 
being in a given quantum state changes in time}. Operationally, it describes the 
\emph{rate at which probability ``flows into or out of'' a chosen state} $\ket{k}$ during the 
system’s evolution. Unlike the standard occupation probability $p_k(t)$, which tells us 
\emph{how likely} the system is to be in $\ket{k}$ at time $t$, the~TF distribution 
$\pi_k(t)$ captures \emph{when} that probability changes most significantly. It can be 
empirically reconstructed by measuring how the population of $\ket{k}$ varies across 
many trials, each measured at a different time, and~taking the magnitude of its 
time~derivative.

We now describe the experimental procedure for reconstructing the TF distribution in a discrete quantum system with eigenstates $\ket{1}, \dots, \ket{n}$ and an observable $\hat{A} = \sum_{j=1}^{n} a_j \ket{j} \bra{j}$.
\begin{enumerate}
    \item \textbf{{Prepare initial state.} }  
    Initialize the system in a fixed quantum state $\hat{\rho}_0$ and~let it evolve unitarily 
    under a given dynamics (open or closed systems).

    \item \textbf{Perform projective measurement at time $t_j$.}  
    At a chosen time $t_j$, perform a projective measurement of the observable $\hat{M}_k = \ket{k}\bra{k}$ to check whether the system is in the target state $\ket{k}$.

    \item \textbf{Repeat to obtain statistics.}  
    Repeat the measurement independently over many trials (typically $N \gg 1$) to estimate the empirical {frequency}  
    $$
    f_k(t_j) = \frac{N_k(t_j)}{N},
    $$  
    where $N_k(t_j)$ is the number of detections in state $\ket{k}$ at time $t_j$. In~the large-$N$ limit, this converges to the probability  
    $
    p_k(t_j) = \mathrm{Tr}(\rho_{t_j} \hat{M}_k) = \bra{k}\rho_{t_j}\ket{k}.
    $

    \item \textbf{Sample over a time grid.}  
    Repeat the above procedure for a discrete set of times $\{t_1, t_2, \dots, t_M\}$ to obtain a time-resolved profile of the population $p_k(t)$. Importantly, each measurement is performed on an independent trial and only once per trajectory. This discrete-time sampling avoids continuous monitoring and thereby prevents Zeno-like inhibition of the dynamics, making it experimentally feasible in discrete systems such as qubit~platforms.

    \item \textbf{Estimate rate of change.}  
    Compute the finite differences  
    $$
    |\Delta f_k(t_j)| = |f_k(t_{j+1}) - f_k(t_j)|,
    $$  
    and define the empirical rate 
    $$
    \tilde{f}_k(t_j) = \frac{|\Delta f_k(t_j)|}{\delta t},
    $$  and define the normalized distribution  
$$
\widehat{\pi }_{k}(t_{j})  = \mathcal{N} \cdot \frac{|\Delta f_k(t_j)|}{\delta t},
$$  
where $\delta t = t_{j+1} - t_j$ is the sampling interval and $\mathcal{N}$ is a normalization constant chosen so that the sum over all time bins satisfies  
$$
\sum_{j=1}^{M-1} \tilde{f}_k(t_j) \cdot \delta t = 1.
$$  

This ensures that $\tilde{f}_k(t_j)$ can be interpreted as a properly normalized discrete approximation to the TF probability distribution. 
\end{enumerate}

This empirical distribution can then be used to compute statistical moments such as the 
mean $\mu = \sum_{j=1}^M t_j \cdot\tilde{f_k}(t_j)$, and~any other higher-order 
momenta as $\mu^{(p)}=\sum_{j=1}^M t_j^p \cdot\tilde{f_k}(t_j)$, as~well as variance in 
the time of flow. 
To construct the continuous-time TF probability distribution from the discrete estimate $\widehat{\pi }_{k}(t_{j})$, 
we take the joint limit where both the number of time steps $M$ and the number of measurement repetitions $N$ tend to infinity:
\begin{equation}\label{Eq:DiscreteTF}
\widehat{\pi }_{k}(t_{j})  \underset{ M,N\rightarrow \infty}{\longrightarrow }\mathcal{N}\left|\frac{d}{dt}p_{k}\left(
t\right)\right| \equiv \pi _k(t)
\end{equation}
where $\mathcal{N}$ is a normalization factor and $ \pi _k(t)$ is the exact continuous-time~distribution.

This analysis suggests that to obtain an analytical exact expression for the TF distribution and predict the experimental data obtained from the experiment described above, one simply needs to calculate the time derivative of the probability for the system to be detected in the state $\ket{k}$
\begin{equation}\label{Eq:TFdistribution}
    \pi_k(t) = \mathcal{N}\left|\frac{d}{dt}p_k(t)\right| = \mathcal{N}\left|\frac{d}{dt}\text{Tr}\left(\widehat{\rho}_t \widehat{M}_k\right)\right| ,
\end{equation}
where $\mathcal{N}$ is the normalization factor (over time).
If $ p_k(t) $ is monotonic, its derivative can be interpreted as a time-of-arrival (TOA) distribution when increasing, or~a time-of-departure (TOD) distribution when decreasing. For~non-monotonic $p_k(t)$, we split the time domain into increasing (TOA) and decreasing (TOD) regions, defining
\begin{align}\label{Eq:TOA/TODdistribution}
    \pi_k^{\text{TOA}}(t) &= \mathcal{N}_A \frac{d}{dt}p_k(t), \quad \ \ \ \text{for } \frac{d}{dt}p_k(t) > 0 , \\
    \pi_k^{\text{TOD}}(t) &= \mathcal{N}_D \left| \frac{d}{dt}p_k(t) \right|, \quad \text{for } \frac{d}{dt}p_k(t) < 0,
\end{align}
with normalization constants $ \mathcal{N}_A $ and $\mathcal{N}_D$ over the 
respective time regions.  
Note that for the non-monotonous case, the~TF distribution given by Equation \eqref{Eq:TFdistribution} captures the distribution of times where the population in the state $\ket{k}$ changes, without~distinguishing between arrival and departure.
Note that the interpretation of Equations~\eqref{Eq:TFdistribution} and \eqref{Eq:TOA/TODdistribution} as normalized quantum flux is analogous to the time-of-arrival (TOA) distribution in continuous systems (see~\cite{Beau24, Beau24_2}). This connection is further developed in the companion paper~\cite{Beau2025_TEUR}, where we introduce a general concept of the time-of-flow (TF) distribution applicable to both continuous and discrete spectra, and~show that the TF distribution coincides with the TOA distribution for a particle propagating in~space. 

By using the master equation $\frac{d}{dt}\widehat{\rho}_t = \mathcal{L}(\widehat{\rho}_t)$, 
where $\mathcal{L}(\cdot)$ is the Liouvillian superoperator~\cite{Breuer2002,lidar2019lecture}, we can rewrite Equation \eqref{Eq:TFdistribution} in a different form
\begin{equation}\label{Eq:TFdistribution:version2:general}
    \pi_k(t) = \mathcal{N}\left|\text{Tr}\left(\mathcal{L}\left(\widehat{\rho}_t\right) \widehat{M}_k\right)\right| =\mathcal{N}\left|\bra{k}\mathcal{L}(\widehat{\rho}_t) \ket{k}\right|.
\end{equation}
For closed systems, we have $\mathcal{L}(\widehat{\rho}_t) = -\frac{i}{\hbar}[\widehat{H},\widehat{\rho}_t]$, where $\widehat{H}$ is the Hamiltonian of the system, and~we obtain the following interesting alternative expression for the TOA distribution
\begin{equation}\label{Eq:TFdistribution:version2:ClosedSystem}
    \pi_k(t) =\mathcal{N}\left|\text{Tr}\left(\widehat{\rho}_t\ \widehat{\Gamma}_k\right)\right| = \mathcal{N}\left|\frac{1}{\hbar}\text{Tr}\left(\widehat{\rho}_t[\widehat{H},\widehat{M}_k]\right)\right|  ,
\end{equation}
where $\text{Tr}\left(\widehat{\rho}_t\ \widehat{\Gamma}_k\right) = \bra{\psi_t}\widehat{\Gamma}_k\ket{\psi_t}$ is the quantum expectation value of the operator $\widehat{\Gamma}_k\equiv-\frac{i}{\hbar}[\widehat{H},\widehat{M}_k]$ 
is the analog of the current operator for continuous systems~\cite{Beau2025_TEUR}. Hence, if~one designs another protocol to measure the mean value of the operator $\widehat{\Gamma}_k$ at different times $t_1,t_2,\cdots, t_M$, then we can reconstruct the TF distribution of the system. This protocol might offer a more direct measurement of the TF distribution for discrete systems such as two-level spin systems, as~we will see~later.

\section{The Two-Level Spin Transition~Model}
\unskip

\subsection{Theoretical~Results} 

We first consider a two-level spin transition model
\begin{equation}\label{Eq:Hamiltoniansigmax}
    \widehat{H}(t) = \dfrac{\hbar \omega(t)}{2}\widehat{\sigma}_x 
\end{equation}
where $\omega(t)$ is a continuous function of $t\geq 0$. The~unitary operator $\widehat{U}(t) = e^{-\frac{i}{\hbar}\int_{0}^t \widehat{H}(t')dt'} = e^{-i\frac{\Omega(t)}{2}\widehat{\sigma}_x} = \cos\left(\frac{\Omega(t)}{2}\right)-i\sin\left(\frac{\Omega(t)}{2}\right)\widehat{\sigma}_x$, where $\Omega(t) = \int_0^t \omega(t')dt'$, see Appendix \ref{A1}, after~Equation (A4), for~details of the derivation. Assuming an initial state $\ket{0}$, the~solution to the solution to the Schrödinger equation reads $\ket{\psi_t} = \widehat{U}_t\ket{0} = \cos\left(\frac{\Omega(t)}{2}\right)\ket{0}-i\sin\left(\frac{\Omega(t)}{2}\right)\ket{1}$, leading to $p_1(t) \equiv |\braket{1|\psi_t}|^2 = \sin^2\left(\frac{\Omega(t)}{2}\right)$. Hence, from~Equation \eqref{Eq:TFdistribution} the TF distribution from state $\ket{0}$ to state $\ket{1}$ reads
\begin{equation}\label{Eq:pi1:spin}
    \pi_1(t)  = \mathcal{N}\left|\omega(t)\sin\left(\Omega(t)\right)\right| .
\end{equation}
As we mentioned before, this distribution represents the TOA distribution (resp. the TOD distribution) if the derivative of $p_1(t)$ is positive (resp. negative). 
Note that we can also use the Equation \eqref{Eq:TFdistribution:version2:ClosedSystem} to find the expression of the TF distribution considering a general initial state $\ket{\psi_0}$
\begin{equation}\label{Eq:TFdistribution:version2:Xgate}
    \pi_1(t) =  \mathcal{N}\left|\omega(t)\cdot\bra{\psi_t}  \widehat{\sigma}_y \ket{\psi_t}\right| = \mathcal{N}\left|\omega(t)\cdot\bra{\psi_0} \widehat{U}_t^\dagger \widehat{\sigma}_y \widehat{U}_t\ket{\psi_0}\right| , 
\end{equation}
where we used the commutator relation $[\widehat{\sigma}_x,\ket{1}\bra{1}]=i\widehat{\sigma}_y$ and where the second equality was obtained from the Heisenberg representation $\widehat{\rho}_t =\widehat{U}_t \widehat{\rho}_0 \widehat{U}_t^\dagger$. The~first equality in \eqref{Eq:TFdistribution:version2:Xgate} implies that the measurement of the mean value of the operator $\widehat{\sigma}_y$ for $t\geq 0$ multiplied by the factor $\omega(t)/2$ and normalized provides us an estimate of the TF distribution. The~second expression in \eqref{Eq:TFdistribution:version2:Xgate} gives another way to calculate explicitly the TF distribution without taking the derivative of $p_1(t)$. 
To illustrate the TF distribution in a solvable setting, we consider a two-level system with Hamiltonian 
$
\hat{H}(t) = \frac{\hbar \omega_0}{2} \hat{\sigma}_x
$
and initial state 
$
\ket{\psi_0} = \cos\left(\frac{\theta}{2}\right)\ket{0} +  \sin\left(\frac{\theta}{2}\right)\ket{1}.
$
For $ \theta \ne \pi/2 $, the~system evolves nontrivially toward $ \ket{1} $, and~the TOA distribution reads
$
\pi_1(t) = \frac{\omega_0}{2} \sin(\omega_0 t), \quad t \leq t_f = \frac{\pi}{\omega_0}.
$
This yields a mean arrival time $ \langle T_1 \rangle = t_f/2 $ and standard deviation 
$
\Delta T_1 = \frac{t_f}{2} \sqrt{1 - \frac{8}{\pi^2}}$. 

\begin{figure}
    \includegraphics[width=1\linewidth]{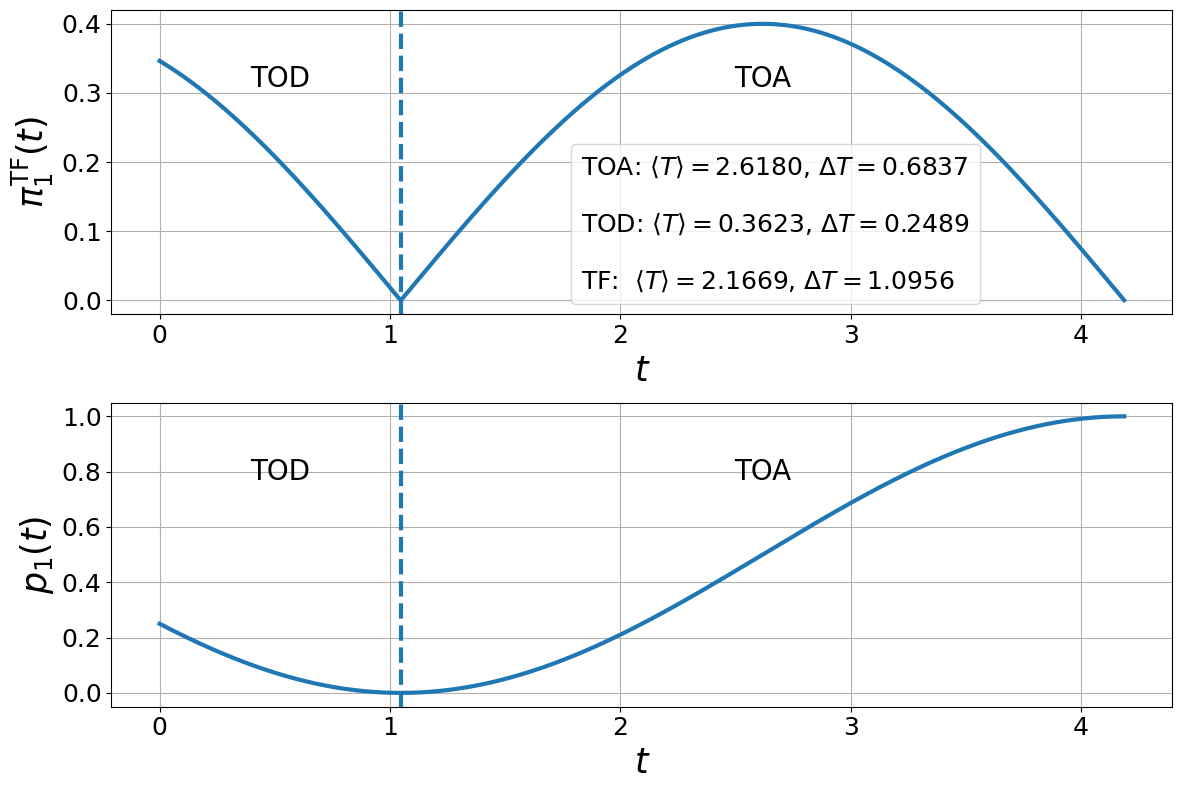}
    \caption{\textbf{Visual representation of TOA and TOD in an oscillating flow.} We plot the TF distribution $\pi(t)$ (top) and probability of occupation in the target state $p_1(t)$ (bottom) 
    for a driven two-level system with constant driving frequency and fixed phases $\theta=\pi/3$ and $\phi=\pi/2$ in \eqref{Eq:InitialState:General}, and~we fixed $\omega_0 =1$. 
    The vertical dashed line at $t \approx 1$ separates the time-of-departure (TOD) regime 
    from the time-of-arrival (TOA) regime, corresponding to the change of sign in $dp_1/dt$.}
    \label{Fig:TF_TOA_TOD}
\end{figure}

{We can show that for a more general initial state
\begin{equation}\label{Eq:InitialState:General}
\ket{\psi_0} = \cos\left(\frac{\theta}{2}\right)\ket{0} + e^{i\phi} \sin\left(\frac{\theta}{2}\right)\ket{1},
\end{equation}
where \( \theta \in [0,\pi] \), \( \phi \in [0, 2\pi) \), and~\( \omega(t) \) is a smooth real-valued function, the~TF distribution reads 
\begin{equation}\label{Eq:TFdistribution:version2:Xgate:General}
\pi_1(t) =\mathcal{N}\cdot \omega(t) \left[\cos(\theta) \sin(\Omega(t)) - \sin(\theta) \cos(\Omega(t)) \sin(\phi)\right], 
\end{equation}
see details of the derivation in Appendix \ref{A1}. Interestingly, when $\theta=\pi/3$ and $\phi=\pi/2$, see Figure~\ref{Fig:TF_TOA_TOD}, we see a transition between two phases: the TOD phase where the population in the target state decreases and the TOA phase where the population of the target state increases up to its maximal value.}

\subsection{Numerical~Optimization} 

\begin{figure}    
    \includegraphics[width=1\linewidth]{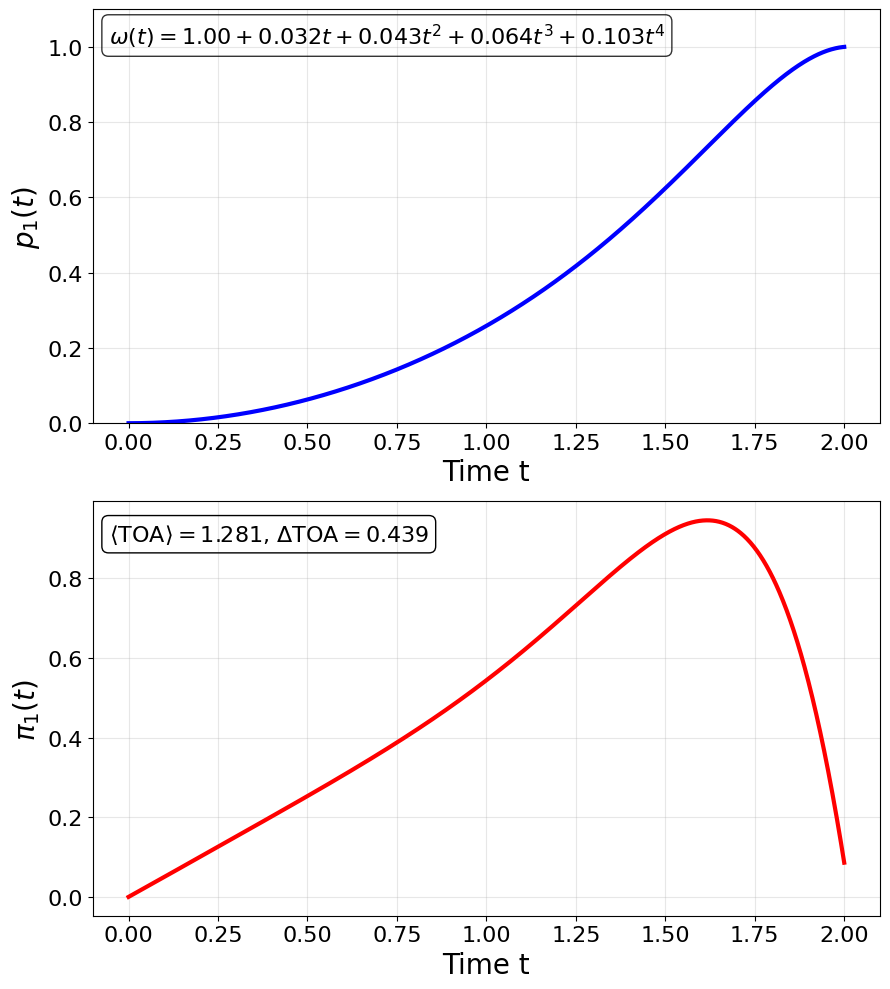}
    \caption{\textbf{Time-dependent control of a two-level quantum system using a polynomial protocol 
\(\omega(t)\).} We optimize the frequency \(\omega(t) = \omega_0 + \sum_{p=1}^{4} 
a_p t^p\) to maximize population transfer from \(|0\rangle\) to \(|1\rangle\) at time 
\(T\), while enforcing monotonic growth of \(p_1(t)\) and regularity of the control. 
The~cost function penalizes non-monotonic behavior and large coefficients. The~optimal 
protocol yields a smooth population trajectory \(p_1(t)\) and the corresponding 
time-of-flow (TF) distribution \(\pi_1(t)\), enabling precise shaping of detection statistics 
through analytically tractable control~functions.}
    \label{Fig:Optimization:sigmax}
\end{figure}

In Figure~\ref{Fig:Optimization:sigmax}, we optimize a time-dependent control protocol $\omega(t)$ driving a two-level quantum system governed by the Hamiltonian \eqref{Eq:Hamiltoniansigmax}. 
Starting from the initial state $\ket{0}$, our goal is to deterministically reach the target state $\ket{1}$ at time $T$, while ensuring that the transition probability $p_1(t)$ increases monotonically over time. We construct $\omega(t)$ as a fourth-order polynomial $\omega(t)=\omega_0+\sum_{p=1}^4 a_p\ t^p$ and define a cost function
\[
J[a_1,a_2,a_3,a_4] = \left(p_1(T) - 1\right)^2 + \lambda_{\mathrm{mono}} \cdot N_{\mathrm{false}} + \lambda_{\mathrm{reg}} \sum_{i=1}^{4} a_i^2,
\]
where $p_1(T)$ is the final population of the target state, $N_{\mathrm{false}}$ counts the number of time points where $dp_1/dt \leq 0$, and~the last term regularizes the control coefficients. The~parameters $\lambda_{\mathrm{mono}}$ and $\lambda_{\mathrm{reg}}$ are penalty weights for monotonicity and regularity, respectively. The~optimization exploits the exact analytical solution of the time-evolved state and provides both the population dynamics and the corresponding TF-distribution $\pi_1(t)$, enabling direct control over temporal detection statistics. This framework offers a versatile and computationally efficient approach to shaping quantum arrival dynamics using smooth, experimentally feasible~controls.

\subsection{The Two-Level Delta-Pulse Model: A Limiting~Case} 

To verify the consistency of our approach, we consider the idealized delta-pulse model
\begin{equation}\label{Eq:deltapulseModel}
\widehat{H}(t) = \frac{\pi\hbar}{2}\widehat{\sigma}_x\delta(t-t_0), 
\end{equation}
describing an instantaneous transition from $\ket{0}$ to $\ket{1}$ at time $t_0$. This model can be obtained from Equations \eqref{Eq:Hamiltoniansigmax} and \eqref{Eq:pi1:spin} with 
$\omega(t)= \pi \cdot e^{-(t-t_0)^2/2\sigma^2}/\sqrt{2\pi\sigma^2}$ 
in the limit $\lim_{\sigma\rightarrow 0} \omega(t) = \pi\delta(t-t_0)$. 
In this case, the~population becomes 
$p_1(t) = \theta(t-t_0)$,
leading to the TOA distribution 
$
\pi_1(t) = \delta(t-t_0),
$ 
and mean TOA $\braket{T_1} =  t_0$, as~expected. The~results obtained in this idealized, 
classical-like limit support the use of the terminology ``time of arrival''. Indeed, satisfying 
this consistency condition, i.e.,~recovering a Dirac delta distribution centered at the 
switching time, is a minimal requirement for any approach claiming to define a TOA 
distribution. Other definitions should be able to reproduce this limit to justify the use of 
such terminology. Further delta-pulse models are analyzed in Appendix \ref{A2}.

\section{Optimization of Shortcut to Adiabaticity (STA) Parameters Using TOA Distributions}
\unskip

\subsection*{STA Model and TOA~Distribution}

\begin{figure}
\includegraphics[width=1.0\linewidth]{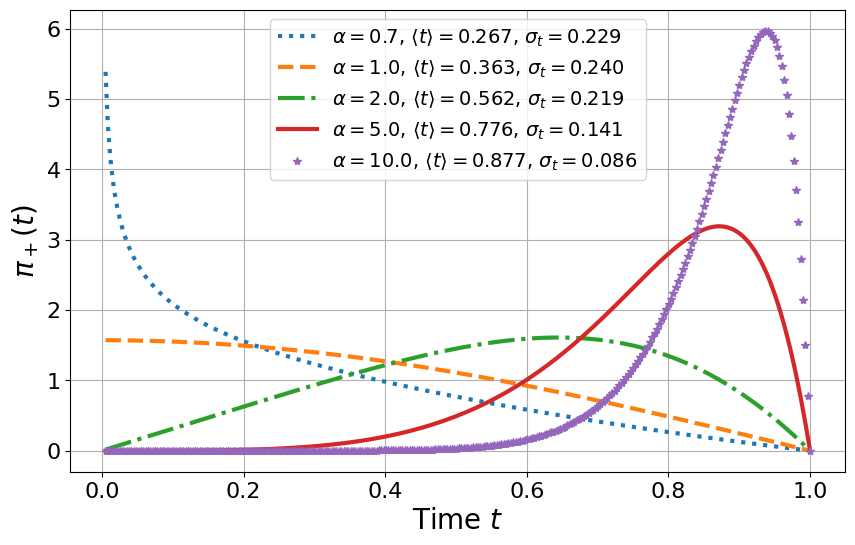}
    \caption{\textbf{Normalized TOA distribution for different $\alpha$ values.} In this figure, we plot the TOA distribution \eqref{Eq:TOAdistribution:STA} for different values of $\alpha$ and the respective mean/standard deviation TOA (see legend within the panel) for $T=1$. }
    \label{Fig:STA:TOAdistrib}
\end{figure}

Let us consider a generalized spin-STA Hamiltonian~\cite{Berry2009,Chen10}
\begin{equation}\label{Eq:STAHamiltonian}
\widehat{H}_{\text{STA}}(t)=\frac{\hbar}{2}\left[-\omega_0(\sin\left(\theta(t)\right)\widehat{\sigma}_x+\cos\left(\theta(t)\right)\widehat{\sigma}_z)+\dot{\theta}(t)\widehat{\sigma}_y\right],
\end{equation}
with a flexible angle parameterization
\begin{equation}\label{Eq:ThetaModel}
\theta(t)=\frac{\pi}{2}\left(\frac{t}{T}\right)^\alpha,\ \alpha\geq 0
\end{equation}
so that the target state is $\ket{+} = \frac{1}{\sqrt{2}}(\ket{0}+\ket{1})$. By~construction of the Hamiltonian, we know the exact solution to the time-dependent Schrödinger equation $\ket{\psi_t} = e^{i\phi_t}\ket{n_t}$, where the Berry phase is $\phi_t=\frac{\omega_0}{2}\int_{0}^t\ dt'\sqrt{1+\frac{\dot{\theta}(t')^2}{\omega_0^2}} $ and where $\ket{n_t} =  \cos\left(\frac{\theta(t)}{2}\right)\ket{0} + \sin\left(\frac{\theta(t)}{2}\right)\ket{1}$ is the solution to the stationary Schrödinger equation $H_0(t)\ket{n(t)} = E_0(t)\ket{n(t)}$, with~$E_0(t) =\mp\frac{\hbar\omega_0}{2} $. From~the overlap between the current state $\ket{\psi_t}$ and the target state $\ket{+}$, we find the probability of occupation of the target state to be $p_{+}(t)\equiv |\braket{+|\psi_t}|^2 = \cos^2\left(\frac{\theta(t)}{2}-\frac{\pi}{4}\right)$, leading to the following expression for the normalized TF distribution
\begin{equation}\label{Eq:TOAdistribution:STA}
    \pi_{+}(t) = \frac{\pi \alpha}{2T}\left(\frac{t}{T}\right)^{\alpha-1}\cos\left(\frac{\pi}{2}\left(\frac{t}{T}\right)^{\alpha}\right),\ 0\leq t \leq T\ ,
\end{equation}
which can be regarded as a TOA distribution since the analysis is restricted to a time interval where the occupation probability increases monotonically.
For the linear model where $\alpha=1$ (see dashed line in Figure~\ref{Fig:STA:TOAdistrib}) that is commonly used experimentally~\cite{Chen10}, we obtain the exact expression 
$
\braket{T_+} = T\left(1-\dfrac{2}{\pi}\right)\approx 0.363\ T$ 
which shows a discrepancy with the target STA time $T$ as~well as an important spread 
given by
$
\Delta T_+ = T\sqrt{\dfrac{4}{\pi}-\dfrac{12}{\pi^2}}\approx 0.240\ T.
$

In Figure~\ref{Fig:STA:TOAdistrib}, we show that different $ \alpha $ values lead to distinct scenarios: (i) mean TOA close to the target time $ T $ with relatively small standard deviation for $ \alpha = 5 $ and $ \alpha = 10 $; (ii) mean TOA closer to $ t = 0 $ with relatively large standard deviation for $ \alpha = 0.7 $; and (iii) intermediate cases for $\alpha = 1 $ and $ \alpha = 2 $. In~this context, the~optimal scenario is obtained for $\alpha = 10$, as~it minimizes both the uncertainty and the difference between the target time $T$ and the mean TOA, and~is therefore more reliable. However, as~the standard deviation of the TOA distribution decreases, the~system becomes more non-adiabatic, which can introduce perturbations and noise. As~a tradeoff, it may be preferable to choose a lower value of $ \alpha$, even at the cost of a larger standard deviation in the TOA distribution to make it~smoother.

\unskip

\section{Three-Level $\Lambda$ Model with Time-Dependent Detuning and Temporal~Magnifier}

\begin{figure*}
\includegraphics[width=0.45\linewidth]{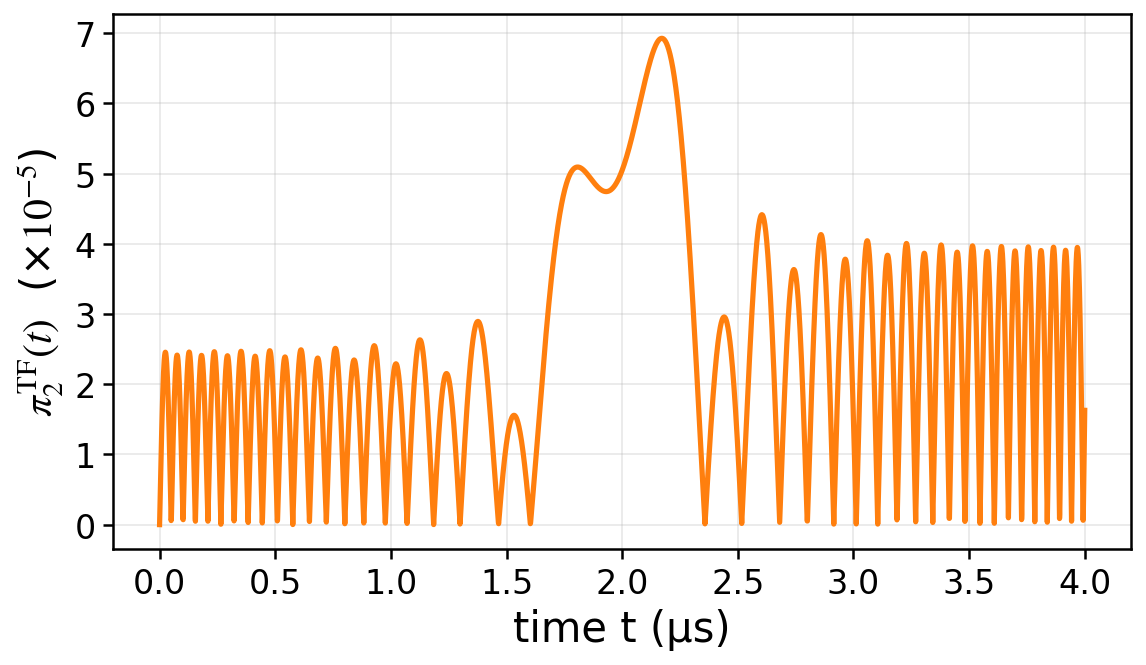}    \includegraphics[width=0.45\linewidth]{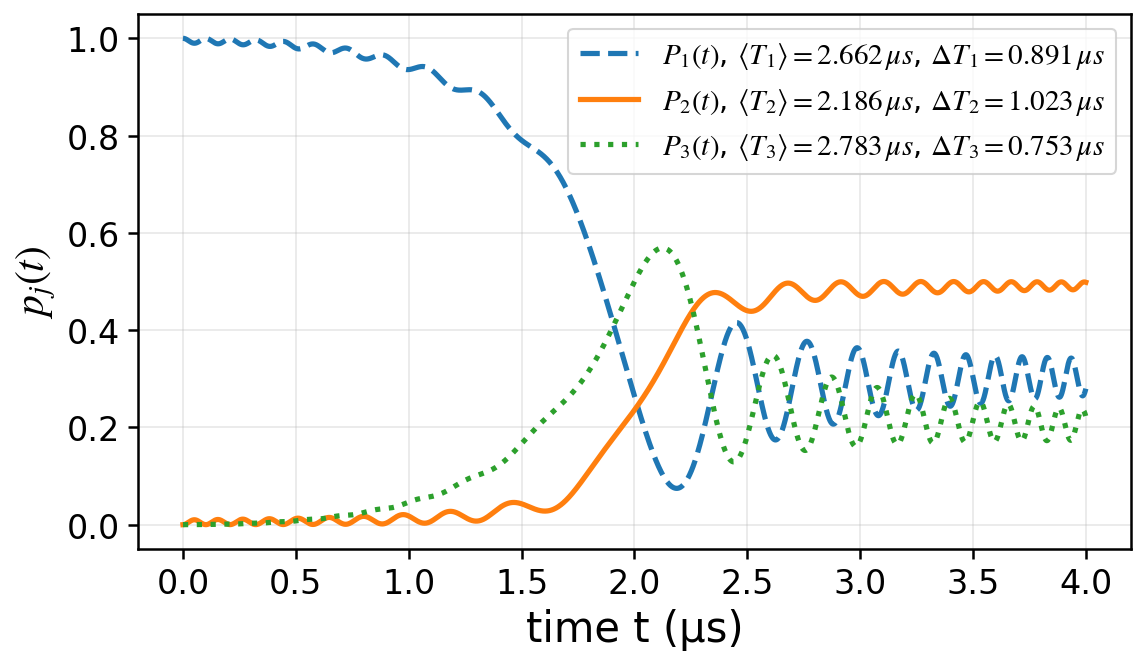}
    \caption{\textbf{Three-level $\Lambda$ model with linear detuning.} In the bottom panel, we represent the time evolution of the populations $p_1(t),\ p_2(t),\ p_3(t)$ for a linear detuning $\Delta(t)/2\pi=-10\to+10$ MHz with constant couplings $\Omega_1=\Omega_2=1$ MHz. The~legend includes TF statistics $\langle T_j\rangle$ and $\Delta T_j$ for each state, illustrating how the TF framework quantifies the characteristic durations of state occupations during the non-adiabatic passage. In~the top panel, we show the normalized TF distribution $\pi^{\rm TF}_2(t)$ (state 2) is obtained from the population flux $d p_2(t)/dt$. }
    \label{Fig:3level}
\end{figure*}

We consider a $\Lambda$ system with ground(-like) states $\{|1\rangle,|3\rangle\}$ and an excited state $|2\rangle$. The~fixed Rabi couplings are $\Omega_1$ on the $|1\rangle \!\leftrightarrow\! |2\rangle$ transition and $\Omega_2$ on the $|3\rangle \!\leftrightarrow\! |2\rangle$ transition. In~the rotating frame and within the rotating-wave approximation, the~Hamiltonian reads
\begin{equation}
\label{eq:LambdaHamiltonian}
H(t)=
\begin{pmatrix}
0 & \tfrac{\hbar\Omega_1}{2} & 0 \\[2pt]
\tfrac{\hbar\Omega_1}{2} & \hbar\Delta(t) & \tfrac{\hbar\Omega_2}{2} \\[2pt]
0 & \tfrac{\hbar\Omega_2}{2} & 0
\end{pmatrix},
\end{equation}
where $\Delta(t)$ is the time-dependent single-photon detuning of the optical field driving $|2\rangle$. With~fixed Rabi couplings $\Omega_{1,2}$ and initial state $|1\rangle$, we sweep the detuning once across resonance, $\Delta_i < 0 < \Delta_f$, implementing the standard $\Lambda$-scheme adiabatic passage~\cite{Bergmann1998,Shore1990}.
We model the detuning as a linear ramp
\begin{equation}
\Delta_{\rm lin}(t) = \Delta_i + \frac{\Delta_f - \Delta_i}{T}\, t \ ,
\end{equation}
with total duration $T$. To~ensure adiabatic following near the avoided crossing (typically around $\Delta = 0$), the~sweep rate is chosen to satisfy a Landau–Zener-type adiabaticity condition $|\dot\Delta| \ll \Omega_{\rm eff}^2$ (as the non-adiabatic transition probability is approximated by the Landau–Zener formula $P_{\rm LZ} = \exp\left(-\pi \Omega_{\rm eff}^2 / 2|\dot\Delta|\right)$), where $|\dot\Delta| = (\Delta_f - \Delta_i)/T$ and $\Omega_{\rm eff} = \sqrt{\Omega_1^2 + \Omega_2^2}$ is the effective coupling between the excited state $\ket{2}$ and the bright state $\ket{B} = (\Omega_1 \ket{3} + \Omega_2 \ket{1})/\Omega_{\rm eff}$. 
In this regime, the~intermediate-state population $p_2(t)$ grows nearly monotonically, peaking within a narrow time window centered on the avoided crossing. This produces a single-lobed TF distribution
$
\pi_2(t) = \mathcal{N}\,\bigl|\tfrac{d}{dt} p_2(t)\bigr|,
$
localized around the crossing point. However, small-amplitude oscillations in $p_2(t)$ lead to significant relative changes in population, which are captured as fine structure in the TF distribution (see Figure~\ref{Fig:3level}). Therefore, the~TF distribution acts as a temporal magnifier, revealing subtle non-adiabatic effects or internal dynamics that are otherwise hidden in population curves. 
While directly reconstructing $\pi_2(t)$ from the derivative of $p_2(t)$ is experimentally challenging, particularly when oscillations make the derivative noisy, we can instead employ the alternative protocol described after Equation~\eqref{Eq:TFdistribution:version2:ClosedSystem}. Computing the current-like operator $\hat{\Gamma}$, one finds
\begin{equation}\label{Eq:Gamma:3level}
\hat{\Gamma} = -\dfrac{\Omega_{\rm eff}}{2}\,\hat{\sigma}_y^{(B,2)} 
\equiv -\dfrac{\Omega_{\rm eff}}{2}\left(-i\ket{B}\bra{2}+i\ket{2}\bra{B}\right)\ ,
\end{equation}
see Appendix \ref{A4} for details. 
Thus, measuring the expectation value $\langle \hat{\Gamma} \rangle$ at different times provides a direct way to reconstruct $\pi_2(t)$, effectively magnifying small non-adiabatic features in the~dynamics.

In Figure~\ref{Fig:3level}, we consider two constant couplings $\Omega_1=\Omega_2=2\pi\times1 $MHz and a linearly swept detuning $\Delta(t)$ from $-10$ to $+10$ MHz over $4\mu$s. Starting from state $\ket{1}$, the~instantaneous dynamics is solved from the time-dependent Schrödinger equation, yielding population transfer between the three states. The~plot on the bottom panel in Figure~\ref{Fig:3level} displays the population dynamics of all three states, $P_1(t),P_2(t),P_3(t)$, with~the corresponding TF mean and standard deviations included in the legend obtained from the respective TF distribution $\pi^{\rm TF}_j(t),\ j=1,2,3$. This illustrates how the TF protocol provides complementary timing information across the whole three-level manifold, beyond~what can be inferred from instantaneous populations alone. The~plot on the top panel in Figure~\ref{Fig:3level} reports the normalized distribution $\pi^{\rm TF}_2(t)$, showing how the population of state 2 is both gained and lost during the transfer. The~fast, high-amplitude oscillations observed in this figure also demonstrate that the transition is rapid and non-adiabatic mentioned~above. 

This application showcases the TF distribution's utility in investigating non-trivial dynamics and opens future avenues to numerically explore protocols that could minimize such oscillations while maximizing population transfer. One suggestion we make is the use of physics-informed neural networks for this task~\cite{Norambuena24}.

\section{Time-of-Flow Distribution and Decoherence in Open~Systems}
\unskip

\subsection{Time-of-Flow Distribution for a General Markovian~System}

Consider the Markovian master equation
\begin{equation}\label{Eq:Master}
    \frac{d}{dt}\hat{\rho}_t = \mathcal{L}(\hat{\rho}_t) = -\frac{i}{\hbar}[\hat{H},\hat{\rho}_t] - \sum_j \frac{\gamma_j}{2}[\hat{L}_j,[\hat{L}_j,\hat{\rho}_t]] , 
\end{equation}
where $\hat{\rho}_t$ is the density matrix, $\hat{H}$ is a time-independent Hamiltonian, $\hat{L}_j$ are time-independent Lindblad operators, and~$\gamma_k$ are the coupling~constants. 

We recall that the TOA distribution at the fixed state $\ket{k}$ is given by
\begin{equation}\label{SM:Eq:TOAdistr}
    \pi_k(t) = \left|\frac{d}{dt}\text{Tr}(\hat{\rho}_t \hat{M}_k)\right| ,
\end{equation}
where $\hat{M}_k = \ket{k}\bra{k}$ is the projector onto the state $\ket{k}$. 

From the master Equation \eqref{Eq:Master}, we find that
\begin{align*}
    \frac{d}{dt}\text{Tr}(\hat{\rho}_t \hat{M}_k) &= -\frac{i}{\hbar}\text{Tr}(\hat{\rho}_t [-\hat{H},\hat{M}]) + \text{Tr}( \hat{\rho}_t\mathcal{D}(\hat{M}_k) ) \\
    &=  \text{Tr}(\rho_t \mathcal{L}^\dagger(\hat{M}_k)),
\end{align*}
where the dissipator $\mathcal{D}(\hat{M}_k) = -\sum_j \frac{\gamma_j}{2}[\hat{L}_j,[\hat{L}_j,\hat{M}_k]] $ and $\mathcal{L}^\dagger(\cdot)$ is the Hermitian conjugate of the Liouville super-operator. 
Hence, we find that
\begin{equation}\label{Eq:TOAdistr:Markovian}
    \pi_k(t) = \mathcal{N}|\text{Tr}(\hat{\rho}_t \mathcal{L}^\dagger(\hat{M}_k))|\ ,
\end{equation}
where $\mathcal{N}$ is the normalization factor. Notice that for a closed quantum system, 
we find the Formula \eqref{Eq:TFdistribution:version2:ClosedSystem}, as~expected. Also, 
the~operator $ \hat{\Gamma} \equiv \mathcal{L}^\dagger(\hat{M}_k) $ can be viewed 
as the analog of a current operator, as~in 
Equation~\eqref{Eq:TFdistribution:version2:ClosedSystem} and~provides a more direct 
way to access the TF distribution, provided that $ \hat{\Gamma} $ is experimentally 
accessible. An~example will be discussed in Section~\ref{Section:Hadamard}.

\subsection{Quantum Speed Limit Associated to the TF~Distribution}

From the master Equation \eqref{Eq:Master}, we find that
\begin{align*}
    \frac{d}{dt}p_k(t) & =  \text{Tr}(\hat{\rho}_t{\mathcal{L}^\dagger}(\hat{M}_k)),
\end{align*}
Hence, we find that
\begin{equation}
   \left|\frac{d}{dt}p_k(t)\right| = |\text{Tr}(\hat{\rho}_t \mathcal{L}^\dagger(\hat{M}_k))| \leq |\sqrt{\text{Tr}( \mathcal{L}^\dagger(\hat{M}_k)^2)|}\ ,
\end{equation}
where we used the Cauchy--Schwarz~inequality. 

Integrating over time from $0$ to $\tau$ leads to
$$
    |p_k(t) - p_k(0)| \leq  \int_0^T |\frac{d}{dt}p_k(t)| \leq T \sqrt{\text{Tr}( \mathcal{L}^\dagger(\hat{M}_k)^2)|}
$$
After parameterizing $p_k(T) - p_k(0) = \theta_T - \theta_0 \equiv \delta \theta$, we find
\begin{equation}\label{Eq:TF-QSL:Open}
    T \geq \tau_{\text{TF}}\equiv\frac{\delta \theta}{\sqrt{|\text{Tr}( \mathcal{L}^\dagger(\hat{M}_k)^2)|}}\ .
\end{equation}
This time-of-flow-like QSL or TF-QSL for transition is the analog to MT-like QSL for open 
quantum systems, see~\cite{delCampo13}. 
Note that for closed system, we obtain the following TF-QSL
\begin{equation}\label{Eq:TF-QSL:Closed}
    \tau_{\text{TF}} = \frac{\hbar \, \delta\theta}{2\Delta_k H}\ ,
\end{equation}
where $\Delta_k H \equiv \sqrt{\mathrm{Tr}(\hat{H}^2\ \hat{M}_k)-\mathrm{Tr}(\hat{H}\ \hat{M}_k)^2}=\sqrt{\bra{k}\hat{H}^2\ket{k}-\bra{k}\hat{H}\ket{k}^2}$ is the standard deviation of the Hamiltonian with respect to the state $\ket{k}$. To~derive the bound \eqref{Eq:TF-QSL:Closed} from the bound \eqref{Eq:TF-QSL:Open}, we use that $|\mathrm{Tr}(\mathcal{L}^\dagger(\hat{M}_k)^2))|=(1/\hbar)|\mathrm{Tr}([\hat{H},\hat{M}_k]^2)|=(1/\hbar)|\mathrm{Tr}(\hat{H}\hat{M}_k\hat{H}\hat{M}_k)-\mathrm{Tr}(\hat{H}^2\hat{M}_k^2)|=(1/\hbar)(\Delta_k H)^2$, as~$\hat{M}_k^2=\hat{M}_k$ and $\mathrm{Tr}(\hat{A}\hat{M}_k)=\bra{k}\hat{A}\ket{k}$. 
This result provides a transition-specific QSL framed in terms of experimentally accessible 
observables via projective measurements at discrete time steps on independently prepared 
systems, avoiding Zeno inhibition. Unlike traditional QSLs based on fidelity or Bures 
angle, this bound focuses on the physical population of a target state and connects 
naturally with experimentally measured quantities, such as the time-of-flow distribution. 
It is particularly well suited for discrete quantum systems, such as qubits, where 
occupation probabilities are the primary observable of~interest.

Other approaches related to observables, such as those based on quantum speed limits 
(QSL) for observables~\cite{GarciaPinto22,Mohan22}, have been previously explored; 
however, these typically emphasize ``what you observe" by bounding the rate of change of 
the expectation value of a Hermitian operator \( A \), using its variance with respect to 
the quantum state. In~contrast, the~time-of-flow (TF) approach addresses the 
complementary question of ``when you observe" by focusing on the dynamics of the 
projective measurement operator \( M_k \), independent of the state. The~resulting 
distribution \( \pi_k(t) = \mathcal{N}|\dot{p}_k(t)| \) characterizes the time statistics of 
detection events, from~which we derive an uncertainty relation and a quantum speed limit 
based on the energy variance \emph{with respect to the measurement operator}. While the 
mathematical tools (e.g., norm inequalities) may be similar, the~TF-QSL differs both 
formally and conceptually from observable-based QSLs by grounding the bound in a 
probabilistic and operational measurement~framework.

\subsection{Lower Bound for the Standard Deviation of the TF Distribution and Uncertainty~Relation}

By Chebyshev's inequality, for~any random variable \( \mathcal{T} \) with distribution \( 
\pi(t) \) and finite variance \( (\Delta \mathcal{T})^2 \), we find the following lower bound: 
\begin{equation}\label{Eq:lowerbound:DeltaTOA}
\Delta \mathcal{T} \geq \frac{1}{3\sqrt{3}}\ \frac{1}{\pi_{\text{max}}} ,
\end{equation}
see~\cite{Beau2025_TEUR} for details of the derivation. 
This inequality shows that the standard deviation \( \Delta t \) of the TF distribution 
cannot be made arbitrarily small if \( \pi_{\text{max}} \) is bounded, enforcing a minimal 
temporal spread inversely proportional to the peak height of the TF distribution. It 
provides a time-uncertainty relation that complements the mean time~analysis.

Now, combining the TF-QSL \eqref{Eq:TF-QSL:Open} with the lower bound of the standard deviation \eqref{Eq:lowerbound:DeltaTOA}, we find
\begin{equation}\label{Eq:lowerbound:DeltaTOA:Final}
\Delta \mathcal{T} \geq \frac{1}{3\sqrt{3}}\ \frac{\delta \theta}{\sqrt{|\text{Tr}( \mathcal{L}^\dagger(\hat{M}_k)^2)|}} = \frac{1}{3\sqrt{3}}\cdot\tau_{\text{TF}},
\end{equation}
where $\delta \theta = p(T)-p(0)$.  
To find this result, it suffices to note that $\pi_{\text{max}}\leq \mathcal{N}\sqrt{|\text{Tr}( \mathcal{L}^\dagger(\hat{M}_k)^2)|}$, as~previously shown, and~that $\mathcal{N}^{-1} = \int_{0}^T \left|\frac{dp_k(t)}{dt}\right|\ dt\geq \left|\int_{0}^T \frac{dp_k(t)}{dt}\ dt\right| = p_k(T)-p_k(0) $. 

Note that using the TF-QSL for closed systems given in Equation \eqref{Eq:TF-QSL:Closed}, we find the following uncertainty relation
\begin{equation}\label{Eq:UncertaintyRelation}
\Delta \mathcal{T}\cdot \Delta_k H \geq \eta\cdot\hbar,
\end{equation}
where $\eta = \dfrac{1}{6\sqrt{3}}\delta\theta=\dfrac{1}{6\sqrt{3}}\left|p_k(T)-p_k(0)\right|$. 

The inequality \eqref{Eq:UncertaintyRelation} provides a novel and operationally grounded uncertainty relation for the TF. This bound emerges from a combination of Chebyshev's inequality and the TF-based quantum speed limit, and~offers a stochastic interpretation based on a clearly defined protocol that measures the TF probability distribution associated with the random variable $\mathcal{T}$. Crucially, it links the temporal spread of the arrival distribution to both the maximal sharpness of the distribution and the energy scale of the evolution, yielding a genuine Heisenberg-like bound. This result shows that temporal resolution in quantum transitions cannot be made arbitrarily small without increasing the energetic resources or sacrificing transition amplitude. As~such, it offers a new tool for assessing the performance and limits of quantum control protocols and time-resolved~measurements. 

Note that in a companion paper~\cite{Beau2025_TEUR}, an~alternative time--energy uncertainty relation is derived, which depends on the standard deviation of the Hamiltonian with respect to the initial state $\rho_0$, defined as 
$\Delta H \equiv \sqrt{\mathrm{Tr}(\hat{H}^2\hat{\rho}_0) - \mathrm{Tr}(\hat{H}\hat{\rho}_0)^2}$. This formulation is particularly useful for Hamiltonians with continuous spectra. The~two uncertainty relations are complementary: the one presented in this article, Equation~\eqref{Eq:UncertaintyRelation}, is applicable to Hamiltonians with discrete spectra and provides a bound for $\Delta \mathcal{T}$ that depends specifically on the energy dispersion within the projector state and offers a direct relation with the QSL as shown in \eqref{Eq:lowerbound:DeltaTOA:Final}, which is applicable for open quantum systems. In~contrast, the~state-dependent bound in~\cite{Beau2025_TEUR}, which extends naturally to both discrete and continuous spectra, provides complementary information on $\Delta \mathcal{T}$, as~it is related to the energy dispersion within the state. Since it depends on the state transfer $\delta\theta$, it is also measurement-dependent. Moreover, this bound is restricted to closed quantum systems, whereas the bound presented here can be naturally extended to open dynamics, as~shown in Equation~\eqref{Eq:lowerbound:DeltaTOA:Final}.

\subsection{Dephasing Model and Quantum Speed~Limits}

We now illustrate the applicability of the TF distribution framework to open quantum systems by considering a paradigmatic pure dephasing model governed by the Lindblad master equation:
\begin{equation}\label{Eq:dephasing:master}
    \frac{d}{dt}\widehat{\rho}_t = -\frac{\gamma}{2}[\widehat{\sigma}_z,[\widehat{\sigma}_z,\widehat{\rho}_t]],
\end{equation}
with initial state \(\widehat{\rho}_0 = \ket{+}\bra{+}\). The~off-diagonal coherence elements in the density matrix decay exponentially, with~a characteristic decoherence time \(\tau_D = (\gamma \Delta \widehat{L}_z^2)^{-1} = 1/\gamma\)~\cite{Chenu17}. Within~our TF distribution framework, the~probability of transition from \(\ket{+}\) to the orthogonal state \(\ket{-}\) evolves as
\begin{equation}
    p_{-}(t) = \frac{1}{2}\left(1 - e^{-2\gamma t}\right),
\end{equation}
yielding the normalized TF distribution (see Equation~\eqref{Eq:TOAdistr:Markovian})
\begin{equation}\label{Eq:dephasing:pim}
    \pi_{-}(t) = \mathcal{N}\left|\frac{d}{dt}p_{-}(t)\right| = 2\gamma e^{-2\gamma t}.
\end{equation}
This distribution is properly normalized, and~provides a well-defined mean arrival time
\begin{equation}
    \braket{T_{-}} = \Delta T_{-} = \frac{1}{2\gamma},
\end{equation}
both equal to half the decoherence time \(\tau_D\). This result demonstrates that the TF distribution captures the temporal scale of decoherence via observable~transitions.

We now evaluate the quantum speed limit (QSL) associated with this transition using the 
TF framework. From~\eqref{Eq:TF-QSL:Open} with the projection operator \(M = 
\ket{-}\bra{-}\) and
$
    \mathcal{L}^\dagger(M) = \gamma(\sigma_z M \sigma_z - M) = \gamma \sigma_x,
$
we obtain
\begin{equation}
    \mathrm{Tr}[\mathcal{L}^\dagger(M)^2] = \gamma^2 \mathrm{Tr}(\sigma_x^2) = 2\gamma^2.
\end{equation}
Since \(p_{-}(0) = 0\) and \(p_{-}(\infty) = \frac{1}{2}\), the~population change is \(\delta \theta = 1/2\). We obtain the TF-based QSL
\begin{equation}
    \tau_{\text{TF}} = \frac{1}{2\sqrt{2}\ \gamma}.
\end{equation}

This result can be directly compared to the fidelity-based Mandelstam–Tamm-type bound derived by del Campo~et~al.~\cite{delCampo13}, which yields
\begin{equation}
    \tau_{\text{MT}} = \frac{1}{\sqrt{2}\ \gamma}.
\end{equation}
Our TF-based QSL is smaller by a factor \(1/2\), which reflects the fundamentally different physical quantities they constrain: MT's bound limits the distinguishability between initial and final states based on fidelity, while our TF-QSL captures the minimal time required to realize a detectable population transfer in a specified state via projective measurements. 
In contrast to global overlap-based speed limits, the~TF-QSL directly links the dynamical generator \(\mathcal{L}\) and the projection operator \(M\), offering a physically transparent and experimentally accessible temporal bound for open system~transitions.

In addition to bounding the mean arrival time, the~TF framework also yields a fundamental lower bound on the standard deviation \(\Delta T_{-}\) of the TF distribution that we can derive from Equation \eqref{Eq:UncertaintyRelation}
\begin{equation}\label{Eq:UncertaintyRelation:DephasingModel}
    \Delta T_{-} \geq \frac{1}{3\sqrt{3}}\cdot \tau_{\text{TF}} = \frac{1}{6\sqrt{6}\ \gamma}.
\end{equation}
This inequality reveals that the temporal spread of arrival times cannot be made arbitrarily small, even in the idealized exponential decay regime. It complements the TF-QSL by capturing a fundamental irreducible fluctuation in the arrival statistics, which is effectively a Heisenberg-type temporal uncertainty rooted in the transition probability and dynamical generator. In~this example, we found \(\Delta T_- = 1/(2\gamma)\), which is about $7.35$ times the limit \mbox{in \eqref{Eq:UncertaintyRelation:DephasingModel}}.

\subsection{Explicit TF--QSL Bound in a Hadamard 
Dephasing~Model}\label{Section:Hadamard}

\begin{figure*}
    \includegraphics[width=0.45\linewidth]{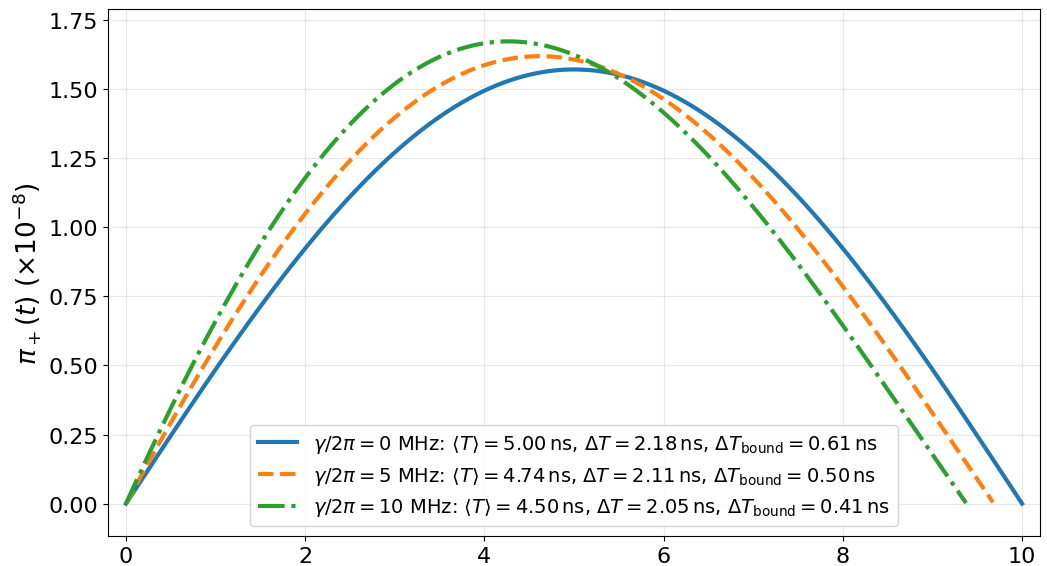}
    \includegraphics[width=0.45\linewidth]{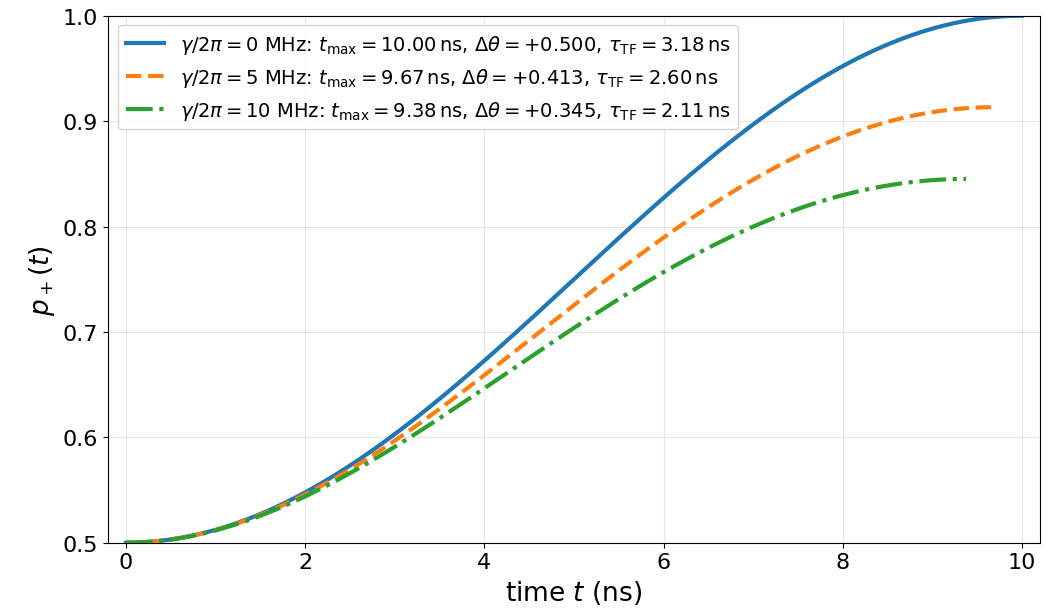}
    \caption{\textbf{TF Distributions and time-uncertainty bound in a Hadamard dephasing.} Normalized TF 
 distributions $\pi(t) = |\mathrm{d}p_+(t)/\mathrm{d}t|$ (in units of $1/\mathrm{ns}$) 
 for increasing dephasing rates $\gamma/2\pi = {0, 5, 10}$ MHz, with~legend reporting 
 $\langle T\rangle$, $\Delta T$, and~the theoretical lower bound $\Delta T_{\rm bound} 
 = \tau_{\rm TF}/(3\sqrt{3})$ is represented on the top panel. On~the bottom panel, we 
 show the population dynamics $p_+(t) = \mathrm{Tr}(\rho_t\ket{+}\bra{+} )$ and 
 effective angular change $\delta\theta$ for each case. The~transition becomes slower and 
 less complete as dephasing increases, validating the predicted scaling of the bound 
 \eqref{Eq:TF-QSL:Hadamard}.}
    \label{Fig:TF:Hadamard:Dephasing}
\end{figure*}

To demonstrate the operational computability of the TF--QSL bound in open dynamics,
we now consider a minimal model consisting of a Hadamard-like rotation with a Markovian dephasing channel along $z$ with rate $\gamma$:
\begin{equation}\label{Eq:MasterEq:Hadamard&dephasing}
    \frac{d}{dt}\hat{\rho}_t=\mathcal{L}(\hat{\rho}_t) = -\frac{i}{\hbar}[\hat{H},\hat{\rho}_t] + \frac{\gamma}{2}
        \big(\hat{\sigma}_z \hat{\rho}_t\hat{\sigma}_z - \hat{\rho}_t\big)  \ ,
\end{equation}
where the Hamiltonian is proportional to the Hadamard gate $\hat{h}=(\hat{\sigma}_x+\hat{\sigma}_z)/\sqrt{2}$
\begin{equation}\label{Eq:MasterEq:Hadamard&dephasing}
    \hat{H} = \frac{\hbar\omega_0}{2}\ \frac{\hat{\sigma}_x+\hat{\sigma}_z}{\sqrt{2}}\ ,
\end{equation}
where the Hadamard gate $\hat{h}$ is transforming the computational basis $\{\ket{0},\ket{1}\}$ to the diagonal basis $\{\ket{+},\ket{-}\}$ from the relations $\hat{h}\ket{0}=\ket{+}$ and $\hat{h}\ket{1}=\ket{-}$.
Hence, the~strategy is to make a transition from the computational to the diagonal basis, thus we consider the initial state $\ket{0}$ and the target state $\ket{+}$. To~realize the protocol, we then use the measurement projector $M_+ = \ket{+}\bra{+}$. 
The goal is to use the protocol to study the dynamics of the transition between the initial and the target states and to find a lower bound for the dispersion of the TF distribution as well as an estimate of the minimal resolution of the detector to analyze this transition. From~the general TF--QSL \eqref{Eq:TF-QSL:Open} and \eqref{Eq:lowerbound:DeltaTOA:Final}, we obtain the explicit lower bound
\begin{equation}\label{Eq:TF-QSL:Hadamard}
   \Delta T_+ \geq \frac{1}{3\sqrt{3}} \tau_{\rm TF} =\frac{1}{3\sqrt{3}}
    \frac{\delta\theta}{\sqrt{\tfrac{\omega_0^2}{4}+\tfrac{\gamma^2}{2}}},
\end{equation}
where $\delta\theta = p_+(T)-p_+(0)$ is the net population transfer; see Appendix 
\ref{A3} for details of the derivation. 
This bound immediately shows how the coherent drive $\omega_0$ and the
dephasing rate $\gamma$ compete in setting the minimal temporal resolution
of the transfer to $|+\rangle$.

To illustrate the behavior of the TF distribution and its associated QSL bound across different decoherence regimes, we numerically solve the open-system dynamics defined by Equation~\eqref{Eq:MasterEq:Hadamard&dephasing}. Figure~\ref{Fig:TF:Hadamard:Dephasing} displays both the normalized time-of-flow (TF) distribution $\pi_+(t)$ and the population dynamics $p_+(t) = \mathrm{Tr}[\rho_t M_+]$ for three representative values of the dephasing rate $\gamma/2\pi = {0, 5, 10}$ MHz. For~each case, we extract the TF distribution from the transition probability curve using $\pi_+(t) = \mathcal{N}|\mathrm{d}p_+(t)/\mathrm{d}t|$, where $\mathcal{N}$ is the normalization factor, compute its mean $\langle T_+ \rangle$ and standard deviation $\Delta T_+$, and~compare the latter to the theoretical bound \eqref{Eq:TF-QSL:Hadamard}. The~lower plot shows how increased dephasing slows down the transition and reduces the effective angular displacement $\delta\theta = p_+(T) - p_+(0)$, thereby enlarging the minimal allowed time uncertainty. This provides a concrete demonstration of the TF--QSL constraint under dissipation, with~the lower bound serving as a resolution benchmark for the detection of population transfer in realistic open quantum~devices. 

Note that the TF distribution $\pi_+(t)$ can be directly obtained by measuring the expectation value of the current-like operator
\begin{equation}\label{Eq:Hadamard}
\hat{\Gamma} \equiv -\dfrac{\omega_0}{2\sqrt{2}}\,\hat{\sigma}_y - \dfrac{\gamma}{2}\,\hat{\sigma}_x
\end{equation}
at different times, see Equation~\eqref{Eq:TOAdistr:Markovian} and Appendix~\ref{A3}. 
Normalizing this quantity over time yields the distribution $\pi_+(t)$, providing a simple and direct route to extracting a time distribution from experimental~data.

\section{Concluding~Remarks} 

In this work, we introduced a general and operationally meaningful definition of the time-of-flow (TF) distribution for discrete quantum systems. The~TF distribution, derived from the rate of change of population in a target state, provides a unifying and experimentally accessible framework to characterize the timing of quantum transitions. In~monotonic regimes, it admits a natural interpretation as a time-of-arrival (TOA) or time-of-departure (TOD) distribution, and~we verified its consistency by recovering the expected delta-function behavior in the limiting delta-pulse~model.

We applied this framework to several scenarios: (i) an analytically solvable two-level 
transition model, including optimization of polynomial control protocols, (ii) optimization 
of smooth shortcut-to-adiabaticity (STA) protocols, (iii) characterization of multi-level 
dynamics through a three-level $\Lambda$ model with time-dependent detuning, (iv) 
analysis of decoherence in open systems using the Lindblad equation, and~(v) analysis of 
TF-based quantum speed limits and uncertainty relations for a Hadamard dephasing 
model. In~each case, the~TF distribution enabled both analytical insight and numerical 
control over the temporal statistics of state~transitions.

Beyond descriptive power, the~TF distribution enabled the derivation of two fundamental results: a quantum speed limit (TF-QSL) specific to population transfer, and~a Heisenberg-like uncertainty relation that bounds the temporal resolution of transitions in terms of the peak height of the TF distribution and the system's dynamical generator. These results connect temporal features of state dynamics to underlying physical resources, offering new tools to assess and optimize quantum control~strategies.

Looking ahead, this framework can guide the design of robust STA protocols by minimizing the mean TOA and its uncertainty while mitigating decoherence. Cost functions combining mean time, standard deviation, and~fidelity offer principled criteria for optimization. The~TF distribution can also serve as a diagnostic of non-adiabatic behavior and as an observable in future experimental studies. Extensions to more multi-level systems, time-dependent noise, and~machine-learning-based optimization in the context of quantum computing are natural next steps for this~approach. 

Looking ahead, the~TF framework could guide the design of robust STA protocols by 
minimizing both mean transition time and its uncertainty and~could serve as a diagnostic 
tool for non-adiabatic dynamics in near-term quantum devices. Extensions to multi-level 
systems, noisy intermediate-scale quantum (NISQ) architectures, 
and~machine-learning-based optimization represent natural next steps. Experimentally, 
the~TF distribution could be reconstructed in superconducting qubits or trapped-ion 
platforms, providing direct benchmarks of control protocols and new probes of temporal 
quantum~uncertainty.\\

\textbf{Acknowledgments.} We thank Lionel Martellini and Lorenzo Maccone for insightful comments. We also thank Simone Roncallo for drawing our attention to the mathematical identity leading to equation \eqref{Eq:TFdistribution:version2:Xgate}.

\bibliography{RefTF}

\newpage
\onecolumngrid

\vspace{5mm} 

\section{APPENDICES}

\subsection{Generalization of the Formula (\eqref{Eq:pi1:spin})}\label{A1}

Consider a two-level system evolving under the time-dependent Hamiltonian
\begin{equation}
\widehat{H}(t) = \frac{\hbar \omega(t)}{2} \widehat{\sigma}_x,
\end{equation}
with initial state
\begin{equation}
\ket{\psi_0} = \cos\left(\frac{\theta}{2}\right)\ket{0} + e^{i\phi} \sin\left(\frac{\theta}{2}\right)\ket{1},
\end{equation}
where \( \theta \in [0,\pi] \), \( \phi \in [0, 2\pi) \), and~\( \omega(t) \) is a smooth real-valued~function.

Let \( \Omega(t) = \int_0^t \omega(t') dt' \). Since \( \widehat{H}(t) \propto 
\widehat{\sigma}_x \), the~time-evolution operator reads
\begin{equation}
\widehat{U}(t) = \exp\left(-\frac{i}{\hbar} \int_0^t \widehat{H}(t') dt'\right) = \exp\left(-i\frac{\Omega(t)}{2} \widehat{\sigma}_x\right).
\end{equation}

First, let us prove the identity
\begin{equation}\label{Eq:Identity:expsigmax}
 e^{-i\phi \sigma_x} = \cos(\phi) \mathbb{I} - i \sin(\phi) \sigma_x 
\end{equation}
using Taylor series 
\begin{align*}
e^{-i\phi \sigma_x} 
&= \sum_{k=0}^\infty \frac{(-i\phi)^k}{k!}\,\sigma_x^k \\
&= \sum_{m=0}^\infty \frac{(-i\phi)^{2m}}{(2m)!}\,\mathbb{I}
  + \sum_{m=0}^\infty \frac{(-i\phi)^{2m+1}}{(2m+1)!}\,\sigma_x.\\
&=\cos(\phi)\,\mathbb{I} - i\sin(\phi)\,\sigma_x
\end{align*}
as 
$$
\sigma_x^{2m} = \mathbb{I}, \qquad 
\sigma_x^{2m+1} = \sigma_x \quad ,\ m \in \mathbb{N}.
$$
and as
\begin{align*}
\sum_{m=0}^\infty \frac{(-i\phi)^{2m}}{(2m)!}
&= \sum_{m=0}^\infty \frac{(-1)^m \phi^{2m}}{(2m)!}
= \cos\phi, \\[6pt]
\sum_{m=0}^\infty \frac{(-i\phi)^{2m+1}}{(2m+1)!}
&= -i \sum_{m=0}^\infty \frac{(-1)^m \phi^{2m+1}}{(2m+1)!}
= -i\sin\phi.
\end{align*}

Now, using the identity \eqref{Eq:Identity:expsigmax} and setting 
$\phi=\dfrac{\Omega(t)}{2} $, we find
\begin{equation}
    \widehat{U}(t) = \cos\left(\dfrac{\Omega(t)}{2}\right)\,\mathbb{I} - i\sin\left(\dfrac{\Omega(t)}{2}\right)\,\sigma_x
\end{equation}

From the expression of the solution to the Schrödinger equation,
\begin{align*}
\ket{\psi(t)} &= \widehat{U}(t) \ket{\psi_0} \\
&= \left(\cos\left(\dfrac{\theta}{2}\right)\cos\left(\dfrac{\Omega(t)}{2}\right)-ie^{i\phi} \sin\left(\dfrac{\theta}{2}\right)\sin\left(\dfrac{\Omega(t)}{2}\right)\right)\ket{0}\\
&+\left(e^{i\phi}\sin\left(\dfrac{\theta}{2}\right)\cos\left(\dfrac{\Omega(t)}{2}\right)-i\cos\left(\dfrac{\theta}{2}\right)\sin\left(\dfrac{\Omega(t)}{2}\right)\right)\ket{1}
\end{align*}
we can compute the probability of finding the system in state \( \ket{1} \):
\begin{align}
p_1(t) & = |\langle 1 | \psi(t) \rangle|^2 \nonumber \\
&= \sin^2\left(\frac{\theta}{2}\right)\cos^2\left(\frac{\Omega(t)}{2}\right) + \cos^2\left(\frac{\theta}{2}\right)\sin^2\left(\frac{\Omega(t)}{2}\right) \nonumber \\
&\quad - \frac{1}{2} \sin(\theta) \sin(\Omega(t)) \sin(\phi).
\end{align}

Differentiating \( p_1(t) \) yields the TF distribution:
\begin{align}
\frac{d}{dt} p_1(t) &= \frac{\omega(t)}{2} \left[\cos(\theta) \sin(\Omega(t)) - \sin(\theta) \cos(\Omega(t)) \sin(\phi)\right], \\
\pi_1(t) &= \mathcal{N} \left| \frac{d}{dt} p_1(t) \right|,
\end{align}
where \( \mathcal{N} \) is the normalization constant for the TF~distribution.

\begin{figure}
    \includegraphics[width=1\linewidth]{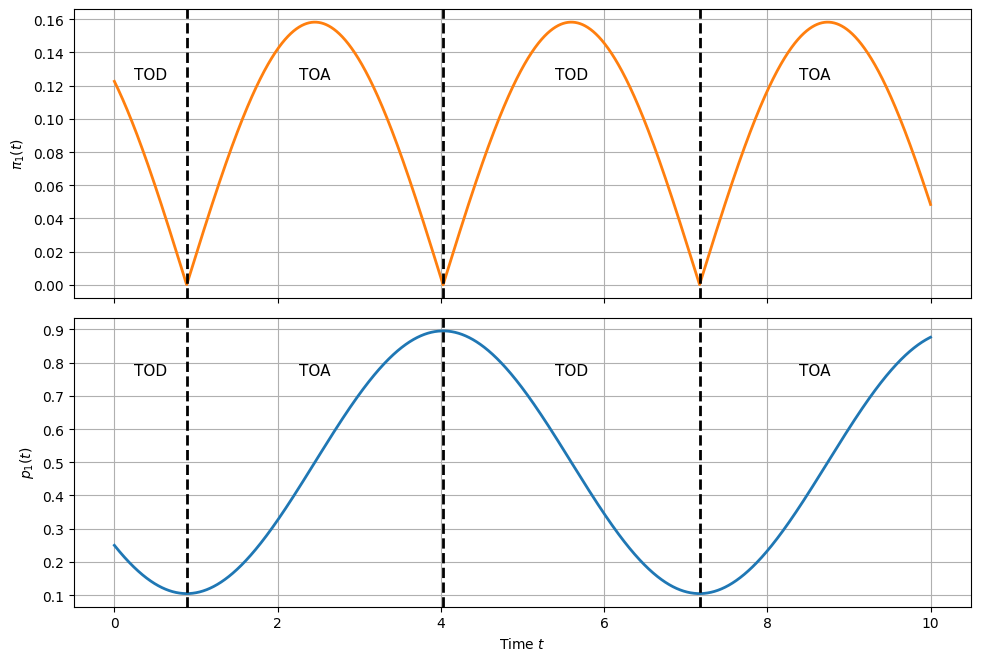}
    \caption{\textbf{Transition dynamics and time-of-flow distribution in a two-level 
    system.} Plots of the state occupation probability \( p_1(t) \) (bottom), 
    the~non-normalized time-of-flow (TF) distribution \( \left| \frac{dp_1}{dt} \right| \) 
    (top), and~the time-of-arrival (TOA) and time-of-departure (TOD) regions for a 
    two-level system with constant Hamiltonian \( \widehat{H}(t) = \frac{\hbar 
    \omega_0}{2} \widehat{\sigma}_x \) and initial state \( \ket{\psi_0} = 
    \cos(\theta/2)\ket{0} + e^{i\phi} \sin(\theta/2)\ket{1} \), with~\( \theta = \pi/3 \) and 
    \( \phi = \pi/4 \). TOA regions correspond to \( \frac{dp_1}{dt} > 0 \), while TOD 
    regions correspond to \( \frac{dp_1}{dt} < 0 \). We fixed $\omega_0 =1$. }
    \label{FigSM:TF_dynamics_two_level}
\end{figure}

When $\phi=0$, we obtain
$$
\frac{d}{dt} p_1(t) = \frac{\omega(t)}{2} \cos(\theta) \sin(\Omega(t)) .
$$
If $\omega(t) = \omega_0$, we find after normalization and restriction in the TOA region ($t\leq \pi/\omega_0$)
$$
 \pi_1(t) = \frac{\omega_0}{2} \sin(\omega_0 t) ,
$$
as shown in the main body of the paper. A~particular case where $\theta=\pi/3$ is shown in Figure~\ref{FigSM:TF_dynamics_two_level}.\\

When $\phi=\pi/2$, we obtain
$$
\frac{d}{dt} p_1(t) = \frac{\omega(t)}{2}\sin(\Omega(t)-\theta) .
$$
So, if~$\omega(t) = \omega_0$, we find that the TF distribution can be interpreted as a TOD distribution for $0\leq t\leq \theta/\omega_0$ and $((2k-1)\pi+\theta)/\omega_0\leq t\leq (2k\pi+\theta)/\omega_0$, see Figure~\ref{FigSM:TF_dynamics_two_level}.

\unskip

\begin{figure}  
    \includegraphics[width=1\linewidth]{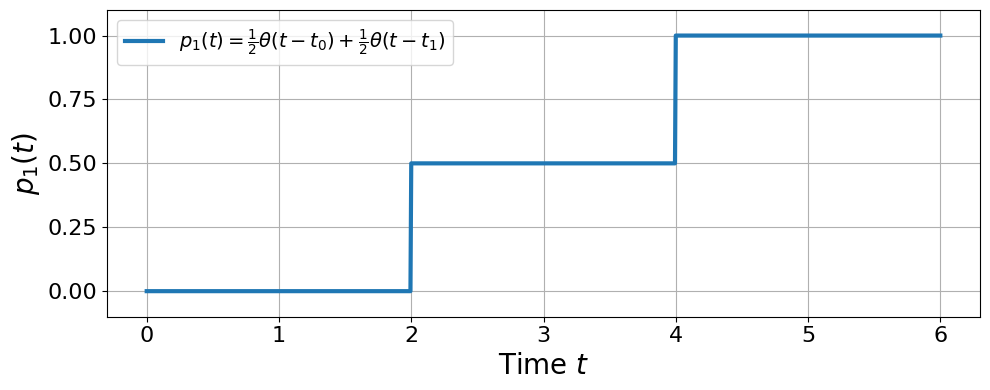}
    \caption{\textbf{Piecewise population function $p_1(t)$ showing the arrival in the state 
    $\ket{1}$.} The flow shows two steps corresponding to two TOA regions for $0\leq t< 4$ 
    and $2< t\leq  6$. As~expected, the~mean TOA is $\dfrac{1}{2}(t_0+t_1)=3$ and the 
    standard deviation is $\frac{1}{2}(t_1-t_0)= 1$.}
    \label{FigSM:p1TOA}
\end{figure}
\vspace{-6pt}

\begin{figure}
    \includegraphics[width=1\linewidth]{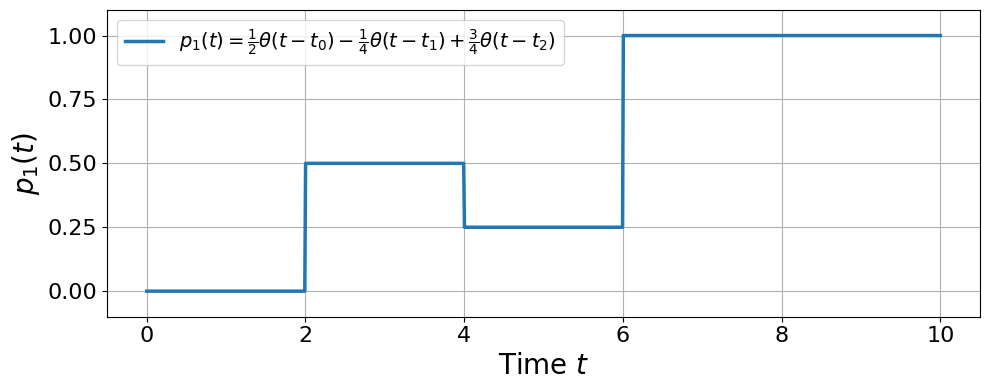}
    \caption{\textbf{Piecewise population function $p_1(t)$ with TOA and TOD.} The flow 
    alternates between TOA regions for $0\leq t< 4$ and $4< t\leq  10$, and~a TOD region 
    for $2< t< 6$. }
    \label{FigSM:p12regions}
\end{figure}
\unskip

\subsection{Limiting Case: The Delta-Pulse~Model}\label{A2}

We can construct a similar model to fill the state $\ket{1}$ in two steps, e.g.,~$p_1(t) = \dfrac{1}{2}\left(\theta(t-t_0)+\theta(t-t_1)\right)$,
which means that half of the spins are expected to be observed in the state $\ket{1}$ between $t_0$ and $t_1$ and the other half after $t_1$. Note that this model can be obtained using a pulse model with a Hadamard gate (operator $\hat{h} = \dfrac{1}{\sqrt{2}}(\hat{\sigma}_x+\hat{\sigma}_z)$) to obtain a $\ket{+} = \frac{1}{\sqrt{2}}(\ket{0}+\ket{1})$ state between $t_0$ and $t_1$, and~then a combined Z-gate and Hadamard gate ($\hat{h}\ \hat{\sigma}_z$ operator) to obtain a $\ket{-}=\dfrac{1}{\sqrt{2}}(\ket{0}-\ket{1})$ and $\ket{1}$ after applying the Z-gate and the Hadamard gate~subsequently. 

Using Equation~\eqref{Eq:TFdistribution}, we now find $\pi_1(t) = \frac{d}{dt}p_{1}(t) = 
\dfrac{1}{2}\delta(t-t_0)+\dfrac{1}{2}\delta(t-t_1)$, whence $\braket{T_1} = 
\dfrac{t_0+t_1}{2}$, which can be interpreted as the TOA distribution for 
$\frac{d}{dt}p_{1}(t)>0$, see Figure~\ref{FigSM:p1TOA}. This means that we expect half of 
the spins to \textit{arrive} in the state $\ket{1}$ at $t_0$ and the other half at $t_1$, 
and~thus, the~average time of arrival should be the average between these two times, 
as~is consistent with the proposed model. Also, the~standard deviation for this two-step 
model is non-zero, $\Delta T_1 = 
\sqrt{\dfrac{1}{2}t_0^2+\dfrac{1}{2}t_1^2-(\dfrac{1}{2}t_0+\dfrac{1}{2}t_1)^2}= 
\frac{1}{2}(t_1-t_0)$, as~expected.

We could of course generalize the approach to a multi-step model with $p_1(t) = \sum_{l=0}^n a_l \theta(t-t_l)$, where $t_0<t_1<\cdots<t_n$ and where $a_l>0,\ \forall l=1,\dots,n$ can be interpreted as the change of probabilities of occupation of the state $\ket{1}$ between two subsequent times $t_{l-1}$ and $t_l$. Here we have  $a_0+a_1+\cdots+a_n = 1$ and the probabilities of occupations at times $t_l$ are then $\sum_{m=0}^{l}a_m$. Of~course, we can also construct TOD distribution by taking $a_l<0,\ \forall l=1,\dots,n$.  

It is also possible to study non-monotonic cases, where the dynamics alternate between TOA and TOD phases, resulting in an overall positive or negative net flow. For~example, take $p_1(t) = \dfrac{1}{2}\theta(t-t_0)-\dfrac{1}{4}\theta(t-t_1)+\dfrac{3}{4}\theta(t-t_2)$ with $t_0=2,\ t_1=4,\ t_2=6$, see Figure~\ref{FigSM:p12regions}, where the flow alternates between TOA regions for $0\leq t< 4$ and $4< t\leq  10$, and~a TOD region for $2< t< 6$. So, the~normalized TOA distribution is 
$$\pi_1^{\text{(TOA)}} =  \dfrac{2}{5}\delta(t-t_0)+\dfrac{3}{5}\delta(t-t_2) $$ 
and the normalized TOD distribution is 
$$
\pi_1^{\text{(TOD)}} = \delta(t-t_1) .
$$
Hence, the~mean TOA is $\langle T_1^{\text{(TOA)}}\rangle = \dfrac{2}{5}t_0+\dfrac{3}{5}t_2 = \dfrac{22}{5}=4.4$ and the mean TOD is $\langle T_1^{\text{(TOD)}}\rangle = t_1 = 4$. Also, the~standard deviation of the TOA is $\sqrt{\dfrac{2}{5}t_0^2+\dfrac{3}{5}t_2^2-\langle T_1^{\text{(TOA)}}\rangle^2 } = \dfrac{4\sqrt{6}}{5}\approx 1.96$ while the standard deviation of the TOD is~zero. 

Now, we can also find the TF distribution to analyze the flow without separating it between TOA and TOD regions. In~this case, the~normalized TF distribution reads
$$
\pi_1^{\text{(TF)}} =  \dfrac{1}{3}\delta(t-t_0)+\dfrac{1}{6}\delta(t-t_1) +\dfrac{1}{2}\delta(t-t_2) .
$$ 
Hence, we obtain that the mean TF, which quantify the average time-of-flow of the dynamics, is $13/3 \approx 4.333$ and the standard deviation of the TF is $\sqrt{29}/3 \approx 1.795$. 

As we see from above, the~mean and standard deviation we obtained for the TOA, TOD, and~TF are the one expected, showing that our approach is valid in these limiting~cases.

\subsection{Proof of the Compact Form of $\Gamma$ for the Three-Level~Model}\label{A4}

We define
\begin{equation}
\hat{\Gamma} \;\equiv\; \frac{i}{\hbar}\,[\hat{H},\ket{2}\bra{2}],
\end{equation}
where the $\Lambda$-system Hamiltonian in the rotating-wave approximation is
\begin{equation}
\hat{H}(t) \;=\; \frac{\hbar\Omega_1}{2}\big(\ket{1}\bra{2}+\ket{2}\bra{1}\big)
+ \hbar\Delta(t)\ket{2}\bra{2}
+ \frac{\hbar\Omega_2}{2}\big(\ket{3}\bra{2}+\ket{2}\bra{3}\big).
\end{equation}
Since the detuning term commutes with $\ket{2}\bra{2}$, it plays no role in the commutator. 
Using the general identity
\[
\big[\ket{a}\bra{b},\ket{2}\bra{2}\big] = \delta_{b,2}\ket{a}\bra{2} - \delta_{a,2}\ket{2}\bra{b},
\]
we obtain
\begin{align}
\hat{\Gamma} 
&= -\frac{\Omega_1}{2}\left(-i\,\ket{1}\bra{2}+i\,\ket{2}\bra{1}\right)
 - \frac{\Omega_2}{2}\left(-i\,\ket{3}\bra{2}+i\,\ket{2}\bra{3}\right) \nonumber \\
&= -\frac{\Omega_1}{2}\,\hat{\sigma}^{(1,2)}_y \;-\; \frac{\Omega_2}{2}\,\hat{\sigma}^{(3,2)}_y ,
\end{align}
where we introduced the Hermitian operators
\[
\hat{\sigma}^{(a,b)}_y \;\equiv\; -i\ket{a}\bra{b} + i\ket{b}\bra{a}.
\]

Next, we define the \emph{bright} and \emph{dark} states
\begin{equation}
\ket{B} = \frac{\Omega_1\ket{1}+\Omega_2\ket{3}}{\Omega_{\rm eff}}, 
\qquad
\ket{D} = \frac{\Omega_2\ket{1}-\Omega_1\ket{3}}{\Omega_{\rm eff}}, 
\qquad
\Omega_{\rm eff}=\sqrt{\Omega_1^2+\Omega_2^2}.
\end{equation}
In this basis one finds
\begin{equation}
\hat{\sigma}^{(B,2)}_y \;=\; -i\ket{B}\bra{2}+i\ket{2}\bra{B}
= \frac{\Omega_1}{\Omega_{\rm eff}}\hat{\sigma}^{(1,2)}_y + \frac{\Omega_2}{\Omega_{\rm eff}}\hat{\sigma}^{(3,2)}_y .
\end{equation}
It follows immediately that
\begin{equation}
\hat{\Gamma} = -\frac{\Omega_{\rm eff}}{2}\,\hat{\sigma}^{(B,2)}_y. 
\end{equation}

Thus, the~current-like operator $\Gamma$ is Hermitian and constant in time 
(for fixed couplings $\Omega_1,\Omega_2$), acts only on the $\{\ket{B},\ket{2}\}$ subspace, 
and leaves the dark state $\ket{D}$ decoupled.

\subsection{TF--QSL Bound for the Hadamard Dephasing~Model}\label{A3}

We have
\begin{align}
    \mathcal{L}^\dagger(\hat{M}_+) &= \frac{i}{\hbar}[\hat{H},\hat{M}_+] + \frac{\gamma}{2}
        \big(\hat{\sigma}_z \hat{M}_+\hat{\sigma}_z - \hat{M}_+\big) \notag \\
    &= -\frac{\omega_0}{2\sqrt{2}}\hat{\sigma}_y - \frac{\gamma}{2}\hat{\sigma}_x.
\end{align}
Hence,
\begin{equation}
    \big(\mathcal{L}^\dagger(\hat{M}_+)\big)^2 =
    \left(\frac{\omega_0^2}{8}+\frac{\gamma^2}{4}\right)\hat{I},
\end{equation}
as $\{\hat{\sigma}_y,\hat{\sigma}_z\}=0$ and $\hat{\sigma}_x^2=\hat{\sigma}_y^2=\hat{\sigma}_z^2=\hat{I}$.
Hence, we obtain
\begin{equation}\label{Eq:A3:Trace}
    \mathrm{Tr}\left[\big(\mathcal{L}^\dagger(\hat{M}_+)\big)^2\right] =
    \frac{\omega_0^2}{4}+\frac{\gamma^2}{2}.
\end{equation}
Combining Equations \eqref{Eq:A3:Trace}, \eqref{Eq:TF-QSL:Open}, and~\eqref{Eq:lowerbound:DeltaTOA:Final}, we obtain Equation~\eqref{Eq:TF-QSL:Hadamard}.

\end{document}